\documentclass{fdsEA} 
\usepackage{amsfonts, amsmath, amssymb, amsthm}
\usepackage{bm}
\usepackage{paralist}
\usepackage[misc]{ifsym}
\usepackage{epsfig} 
\usepackage{epstopdf} 
\usepackage[colorlinks=true]{hyperref}
\hypersetup{urlcolor=blue, citecolor=red}
\allowdisplaybreaks

\def\currentvolume{X}
 \def\currentissue{X}
  \def\currentyear{200X}
   \def\currentmonth{XX}
    \def\ppages{X-XX}
     \def\DOI{10.3934/fods.2023013}

\newcommand{\p}{\bm}
\newcommand{\red}[1]{{\color{red} #1}}
\newcommand{\blue}[1]{{\color{blue} #1}}
\newcommand{\green}[1]{{\color{green} #1}}

\newcommand{\bq}{\begin{eqnarray*}}
\newcommand{\eq}{\end{eqnarray*}}
\newcommand{\bqn}{\begin{eqnarray}}
\newcommand{\eqn}{\end{eqnarray}}

\newtheorem{theorem}{Theorem}[section]

\theoremstyle{definition}

\title[Dynamic Topological Data Analysis]
{Dynamic topological data analysis of functional human brain networks} 

\author[Moo K. Chung, Soumya Das and Hernando Ombao]{}


\thanks{$^*$Corresponding author: Moo K. Chung}

\begin{document}
\maketitle

\centerline{\scshape
Moo K. Chung$^{{\href{mailto:mkchung@wisc.edu}{\textrm{\Letter}}}*1}$,
Soumya Das$^{{\href{mailto:sundew122436@gmail.com}{\textrm{\Letter}}}1}$
and Hernando Ombao$^{{\href{mailto:hernando.ombao@kaust.edu.sa}{\textrm{\Letter}}}2}$
}

\medskip

{\footnotesize
 \centerline{$^1$University of Wisconsin-Madison, USA}
} 

\medskip

{\footnotesize
 \centerline{$^2$King Abdullah University of Science and Technology, Saudi Arabia}
}

\bigskip

 \centerline{(Communicated by Vasileios Maroulas)}


\begin{abstract}
Developing reliable methods to discriminate different transient brain states that change over time is a key neuroscientific challenge in brain imaging studies. Topological data analysis (TDA), a novel framework based on algebraic topology, can handle such a challenge. However, existing TDA has been somewhat limited to capturing the static summary of dynamically changing brain networks. We propose a novel dynamic-TDA framework that builds persistent homology over a time series of brain networks. We construct a Wasserstein distance based inference procedure to discriminate between time series of networks. The method is applied to the resting-state functional magnetic resonance images of the human brain.
We demonstrate that our proposed dynamic-TDA approach can distinctly discriminate between the topological patterns of male and female brain networks. MATLAB code for implementing this method is available at \url{https://github.com/laplcebeltrami/PH-STAT}.
\end{abstract}


\section{Introduction}

The dynamical changes in brain function and activity are an archetypal example of complex systems. Within this system, brain functions depend on constant interplays between local information processing and efficient global integration of information \cite{sporns.2005, bullmore.2009, siemon.2014}.  To reveal the underpinnings of neurodegenerative diseases, it is critical to analyze such interactions in brain pathology \cite{uhlhaas.2006, uhlhaas.2010, hu.2015}. In brain network analysis, the neural interactions are encoded as a graph consisting of nodes and edges. Often, the whole brain is parcellated into several disjoint regions, which are represented as network nodes \cite{tzourio.2002, desikan.2006, hagmann.2007, fornito.2016, arslan.2018} whereas the correlations between different parcellations (nodes) are represented as edge weights. In most analysis based on graph theory, features such as node degrees and clustering coefficients are acquired from adjacency matrices after thresholding edge weights \cite{sporns.2003, vanwijk.2010}. However, this leads to final statistical results and interpretations that rely on the choice of the threshold \cite{lee.2012.TMI, chung.2013.MICCAI} - which is subjective. Given this limitation, developing a multiscale network model that yields reliable outcomes independent of threshold choice is essential. Topological data analysis (TDA) can fill this gap by providing a topologically consistent solution across varying thresholds \cite{edelsbrunner.2000, zomorodian.2005, singh.2008, ghrist.2008, carlsson.2008, wasserman.2018}. Within TDA, persistent homology based approaches have become increasingly popular as a tool for analyzing different brain imaging data because they can capture the persistences \cite{oballe2021bayesian, song.2020.ISBI} of different topological features that are robust under different scales \cite{bubenik.2007, ghrist.2008, lee.2012.TMI, chung.2019.NN}. The persistences are usually summarized and expressed using barcodes and persistence diagrams \cite{song.2023}. Although such approaches have been applied to increasingly diverse biomedical problems, they are mostly limited to investigating the {\it static} summary of {\it dynamically changing} data such as functional magnetic resonance images (fMRI) and electroencephalography (EEG) \cite{chung.2019.ISBI, bassett.2017, wang.2018.annals}.  A recent development involves the application of TDA to analyze dynamic patterns in various datasets, including financial data \cite{gidea.2018} and gene expression data \cite{perea.2015, song.2020.ISBI}.

Motivated by these studies, we develop a novel dynamic-TDA based approach that can statistically discriminate between two groups of brain networks,  reflecting the dynamic nature of brain functional processes. This dynamic-TDA framework builds persistent homology over a multivariate time series \cite{song.2020.ISBI} and constructs a test statistic using the Wasserstein distance between persistence diagrams. Our proposed dynamic-TDA, when applied to resting-state fMRI (rs-fMRI) data, reveals notable topological differences in the temporal dynamics between male and female subjects. This approach moves beyond the traditional methods that focus on static summaries of dynamic images \cite{chung.2019.ISBI, bassett.2017, wang.2018.annals}, allowing for a more nuanced exploration of dynamic patterns within the images \cite{gidea.2018, perea.2015, song.2020.ISBI}. Motivated by time-delay embedding (TDE), we introduce a novel Dynamic Embedding technique to capture the temporal evolution of topological changes in multivariate time series data \cite{von.2010.tde}

\section{Methods}

Human brain functional networks can be modeled as graphs, with nodes representing parcellated brain regions and edges indicating interactions between these regions \cite{chung.2019.NN}. We construct graph filtrations on these networks as described below. During this filtration process, topological features such as connected components (0-dimensional homology) and cycles (1-dimensional homology) emerge and vanish. A feature that appears at a filtration value \( b_i \) and disappears at \( d_i \) is represented as a point \( (b_i, d_i) \) in the plane. The set of all such points forms the persistent diagram (PD), which encapsulates the topology of the underlying network. When these features are represented as a collection of intervals on the real line, they constitute barcodes \cite{carlsson.2009.bulletin}.


\subsection{Graph filtration}

Let $G = (V, W)$ be a weighted graph, where $V$ is a set of nodes and $W$ is the set of edge weights. The edge weight $w_{ij}$ between nodes $i$ and $j$ are assumed to be unique. We assume there are $|V| = p$ number of nodes and $|W| = q = p(p-1)/2$ number of edges.  A binary graph $G_\epsilon = (V, W_\epsilon)$ of $G$ is a graph with binary edge weight $w_{ij,\epsilon}$:
\begin{align*}
w_{ij, \epsilon} = \begin{cases}
1, \text{~~~if~} w_{ij}>\epsilon,\\
0,  \text{~~~otherwise.}
\end{cases}
\end{align*}

A graph filtration of $G$ is defined as a collection of nested binary networks:
\begin{align*}
G_{\epsilon_0} \supset G_{\epsilon_1} \supset \cdots \supset G_{\epsilon_k},
\end{align*} where $\epsilon_0 <\epsilon_1<\cdots<\epsilon_k$ are filtration values \cite{anand.2023.TMI,chung.2019.NN, song.2023}.

By arranging the edge weights in increasing order,
\begin{equation}
w_{(1)} = \min_{j,k} w_{jk} < w_{(2)} < \cdots < w_{(q)} = \max_{j,k} w_{jk},
\end{equation}
where the subscript \(_{(\;)}\) denotes the order statistic, we construct the graph filtration at these edge weights \cite{chung.2019.NN,song.2023}:
\begin{equation}
G_{w_{(1)}} \supset G_{w_{(2)}} \supset \cdots \supset G_{w_{(q)}}.
\end{equation}

\begin{figure}[t]
	\centering
	\includegraphics[width=1\linewidth]{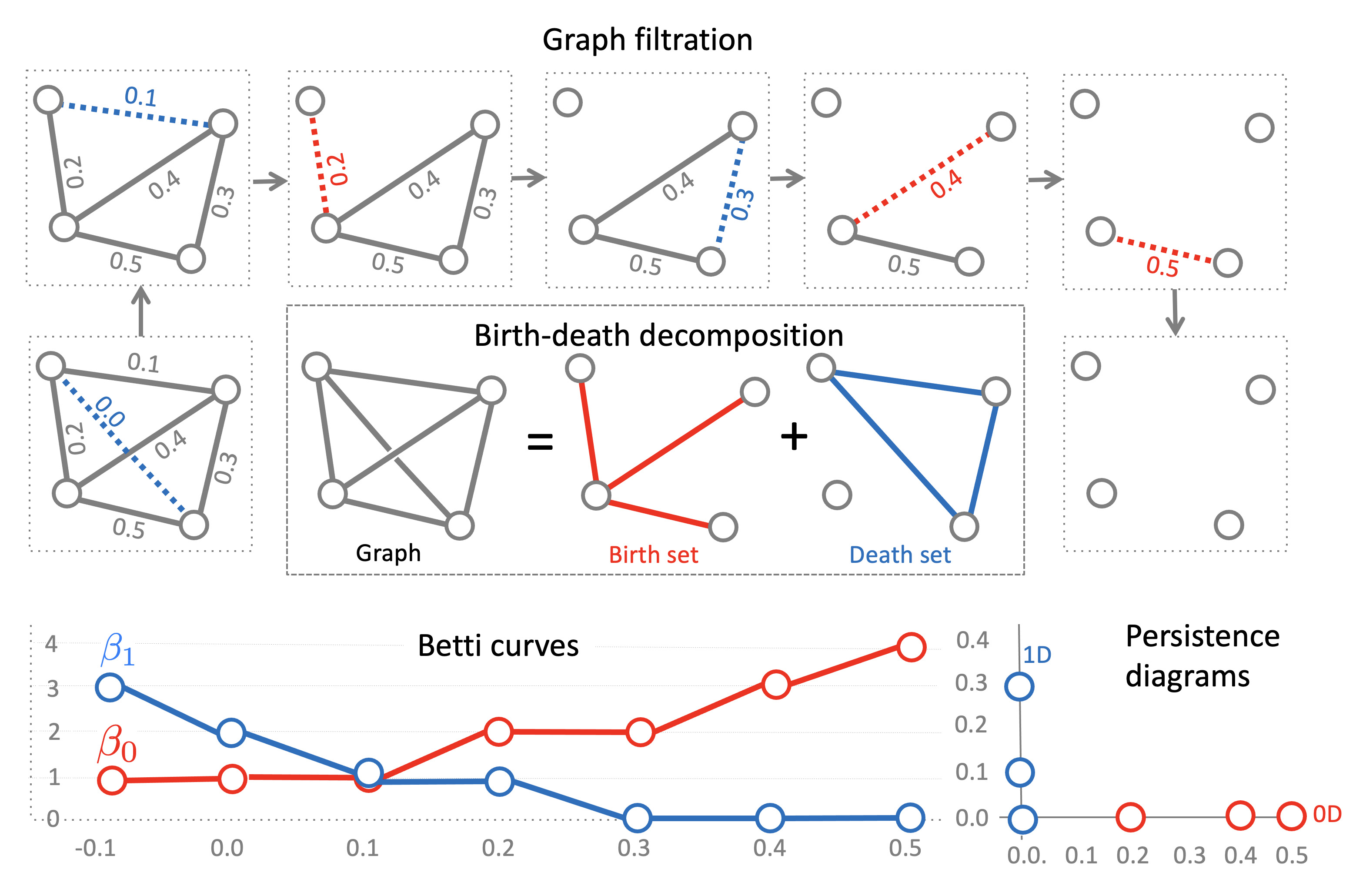}
	\caption{Graph filtration and the corresponding birth-death decomposition of a graph with $4$ nodes. The birth-death decomposition partitions the edge set into disjoint birth and death sets. The resulting Betti curves are monotonic while the persistence diagrams are 1-dimensional points in $\mathbb{R}$.}
\label{fig:bd_decomposition}
\end{figure}

The graph filtration of a graph with $4$ nodes is illustrated in Figure~\ref{fig:bd_decomposition}. Unlike the Vietoris-Rips filtration \cite{carlsson2022applying, zhangpersistent}, the graph filtration does not produce more than 1-dimensional homology.   In a binary network $G_\epsilon$, which is a simplicial complex consisting of only nodes and edges, $0$-dimensional ($0$D) holes are \textit{connected components} and $1$-dimensional ($1$D) holes are \textit{cycles}.
The number of connected components and the number of independent cycles in a graph are referred to as the \(0^{\text{th}}\) Betti number \((\beta_0)\) and \(1^{\text{st}}\) Betti number \((\beta_1)\), respectively.
In a graph filtration, the sequential removal of edges leads to either the birth of a connected component or the death of a cycle \cite{chung.2019.NN}. Once a connected component is born, it never dies. Therefore its death values are assumed to be at $\infty$ and ignored. On the other hand, all  cycles are considered to be born at $-\infty$ and their birth values  are ignored. Thus we can partition edge weights into the birth set  $B(G)$ and death set $D(G)$. The birth set  $B(G)$ consists of the sequence of increasing birth values
\begin{align*}
B(G): b_1< b_2 < \cdots< b_{m_0},
\end{align*} where $m_0 = p-1$ \cite{song.2023}. Ignoring the death values at $\infty$, 0-dimensional persistent diagram is simply characterized by the birth set $B(G)$ (Figure \ref{fig:bd_decomposition}). Further, the persistence diagram is also equivalent to the barcode. The death set  $D(G)$  consists of the sequence of increasing death values
\begin{align*}
D(G): d_1< d_2 \cdots< d_{m_1}.
\end{align*}

During the graph filtration, the removal of an edge  splits the graph into at most two components.
Therefore, the number of connected components \( \beta_0 \)  increases, and the increase is at most one \cite{chung.2019.NN}.
The Euler characteristic \( \chi \) of the graph is given by \cite{adler.2010},
\[
\chi = p - q = \beta_0 - \beta_1.
\]

Thus,
\[
\beta_1 = \beta_0 - p + q.
\]

Note that \( p \) is fixed over the filtration, but \( q \) decreases by one while \( \beta_0 \) increases by at most one. Hence, \( \beta_1 \) always decreases, and the decrease is at most one \cite{chung.2019.NN}. The monotonic properties of Betti numbers over filtration is illustrated in Figure~\ref{fig:bd_decomposition}. The total number of death values for a complete graph
with $p$ nodes was derived in \cite{song.2023} to be
$$m_1 = \beta_1 = 1 - p + \frac{p(p-1)}{2} = \frac{(p-1)(p-2)}{2}.$$

Ignoring the birth values at $-\infty$, 1-dimensional persistent diagram is characterized by the death set $B(G)$ (Figure \ref{fig:bd_decomposition}).  This also diverges from the Vietoris-Rips filtration, which produces cloud points  in
$\mathbb{R}^2$ as the 1-dimensional persistent digram \cite{ berry.2020,bubenik.2015, chazal.2014,edelsbrunner.2008}.
Every edge of a graph must belong to either the birth set or the death set. This leads to the \textit{birth-death decomposition} of a graph, which partitions the set of edge weights into the birth and death sets \cite{song.2023} as follows
\[
W = B(G) \cup D(G), \quad B(G) \cap D(G) = \emptyset.
\]

The birth-death decomposition of a graph is illustrated in Figure \ref{fig:bd_decomposition}.

\subsection{Wasserstein distance}
The topological similarity or dissimilarity between two networks can be quantified using the \textit{Wasserstein distance} between persistent diagrams \cite{liang.2018, mi.2020,1974.vallender, song.2021.MICCAI}. Let \( G_1 = \left(V, W_1\right) \) and \( G_2 = \left(V, W_2\right) \) be two given networks, each with $p$ nodes. Let \( P_1 \) and \( P_2 \) be the corresponding persistent diagrams. The \( r \)-Wasserstein distance between \( P_1 \) and \( P_2 \) is then given by
\[
\mathfrak{L}_r(P_1, P_2) = \inf_{\tau: P_1 \to P_2} \left( \sum_{x \in P_1} \| x - \tau(x) \|^{r} \right)^{1/r},
\]
where the infimum is taken over every possible permutation or bijection \( \tau \) between \( P_1 \) and \( P_2 \) \cite{berwald2018computing, panaretos.2019}. As \( r \) approaches infinity, the \( \infty \)-Wasserstein distance is  given by:
\[
\mathfrak{L}_\infty(P_1, P_2) = \inf_{\tau: P_1 \to P_2} \max_{x \in P_1} \| x - \tau(x) \|
\]

This distance is also known as the \textit{Bottleneck distance} \cite{cohen.2007.stability}.
	
Standard methods for computing the Wasserstein distance, such as the Kuhn-Munkres and Hungarian algorithms, typically have a computational complexity of \( \mathcal{O}(p^3) \) \cite{edmonds.1972, munkres1957algorithms, song.2023, su2015optimal, su2015shape}. In contrast, for graph filtrations that yield 1-dimensional scatter plots as persistence diagrams, the Wasserstein distances can be calculated more efficiently with a computational complexity of \( \mathcal{O}(p \log p) \) by directly matching sorted birth and death values.

\begin{theorem}\label{thm1}
For persistent diagrams \( P_1 \) and \( P_2 \) from graph filtrations, we have
\[
\mathfrak{L}_r(P_1, P_2) = \inf_{\tau: P_1 \to P_2} \left( \sum_{x \in P_1} |x - \tau(x) |^r \right)^{1/r} = \left( \sum_{x \in P_1} | x - \tau^*(x) |^r \right)^{1/r},
\]
where \( \tau^* \) maps the \( i \)-th smallest value in \( P_1 \) to the \( i \)-th smallest value in \( P_2 \) for all \( i \). For \( r\to \infty \),
\[
\mathfrak{L}_\infty(P_1, P_2) = \inf_{\tau: P_1 \to P_2} \max_{x \in P_1} | x - \tau(x)| = \max_{x \in P_1} |x - \tau^*(x)|.
\]
\end{theorem}

\begin{proof}
We prove for 0-dimensional persistence diagrams only; the proof for 1-dimensional diagrams is analogous. Suppose \(P_1\) and \(P_2\) consist of birth values
\[
P_1: b_1^1 < b_2^1 < \cdots < b_{m_0}^1, \quad P_2: b_1^2 < b_2^2 < \cdots < b_{m_0}^2
\]
respectively. Note $\tau^*$ is the identity permutation, i.e., $\tau^*(i) = i$. We aim to prove that
\[
\sum_{i=1}^{m_0} |b_i^1 - b_i^2|^r \leq \sum_{i=1}^{m_0} |b_i^1 - b_{\tau(i)}^2|^r
\]
for any permutation \( \tau \) that is not the identity.

Consider a non-identity permutation \( \tau \). $\tau$ will include at least one pair \( (i, \tau(i)) \) where \( i \neq \tau(i) \).
Without loss of generality, assume \( i < \tau(i) \). Then, \( b_i^1 < b_{\tau(i)}^1 \) and \( b_i^2 < b_{\tau(i)}^2 \). Hence,
$$b_i^1 - b_i^2 < b_{\tau(i)}^1 - b_i^2, \quad  b_{\tau(i)}^1 - b_{\tau(i)}^2 < b_{\tau(i)}^1 - b_i^2.$$

Subsequently,
\[
|b_i^1 - b_i^2|^r + |b_{\tau(i)}^1 - b_{\tau(i)}^2|^r < |b_i^1 - b_{\tau(i)}^2|^r + |b_{\tau(i)}^1 - b_i^2|^r,
\]

Summing over all such pairs \( (i, \tau(i)) \) shows that
\[
\sum_{i=1}^{m_0} |b_i^1 - b_i^2|^r < \sum_{i=1}^{m_0} |b_i^1 - b_{\tau(i)}^2|^r,
\]
thus proving the first statement. The second statement follows as
$$
\mathfrak{L}_\infty(P_1, P_2) = \lim_{r \to \infty} \mathfrak{L}_r(P_1, P_2) = \lim_{r \to \infty} \left( \sum_{x \in P_1} | x - \tau^*(x) |^r \right)^{1/r} = \max_{x \in P_1} |x - \tau^*(x)|.
$$
\end{proof}

The theorem is related to the majorization theorem \cite{marshall.1979,zaheer.2019}.
Given ordered vectors \( \mathbf{\alpha} = (\alpha_1, \cdots, \alpha_n) \)
and \( \mathbf{\beta} = (\beta_1, \cdots, \beta_n) \) satisfying
\[
\alpha_1 \geq \alpha_2 \geq \cdots \geq \alpha_n, \quad \beta_1 \geq \beta_2 \geq \cdots \geq \beta_n,
\]
\( \mathbf{\alpha} \) majorizes \( \mathbf{\beta} \) if
\[
\sum_{i=1}^{k} \alpha_i \geq \sum_{i=1}^{k} \beta_i, \quad \sum_{i=1}^{n} \alpha_i = \sum_{i=1}^{n} \beta_i.
\]
for all \( 1 \leq k \leq n \). For \( \mathbf{\alpha} \) majorizing \( \mathbf{\beta} \) and for any convex function \( f \), we have \cite{marshall.1979,zaheer.2019}
\[
\sum_{i=1}^n f(\alpha_i) > \sum_{i=1}^n f(\beta_i).
\]

Then by identifying $ \alpha_i$ as the \( i \)-th largest element in the set \( \left\{ |x - \tau(x)| : x \in P_1 \right\} \), \( \beta_i \) as the \( i \)-th largest element in the set \( \left\{ |x - \tau^*(x)| : x \in P_1 \right\} \), and convex function $f (x)= |x|^r$, we obtain the desired result.

\begin{theorem}
The Wasserstein distance satisfies the following inequality
\begin{align*}
\mathfrak{L}_{\infty}(P_1, P_2) \leq \mathfrak{L}_p(P_1, P_2) \leq \mathfrak{L}_r(P_1, P_2) \leq \mathfrak{L}_1(P_1, P_2),
\end{align*}
for \( r < p < \infty \).
\end{theorem}

\begin{proof}
We will restrict our proof to 0-dimensional persistence diagrams; the proof for 1-dimensional diagrams follows analogously.
Let \(P_1\) and \(P_2\) consist of birth values
\[
P_1: b_1^1 < b_2^1 < \cdots < b_{m_0}^1, \quad P_2: b_1^2 < b_2^2 < \cdots < b_{m_0}^2
\]
respectively. The \(r\)-Wasserstein distance between connected components is given by
\[
\mathfrak{L}_r(P_1, P_2)=  \sum_{i=1}^{m_0}  ( b_{i}^1 - b_{i}^2 )^2.
\]
Note that the sorted vector $(b_1^k, b_2^k, \cdots, b_{m_0}^k)^{\top}$ is 
a point in the \((m_0-1)\)-simplex \(\mathcal{T}_0\) defined as
\[
\mathcal{T}_0 = \{ (x_1, x_2, \cdots, x_{m_0})  | x_1 < x_2 < \cdots < x_{m_0} \} \subset \mathbb{R}^{m_0}
\]
with $x_1$ bounded below and $x_{m_0}$ bounded above. Hence, the 0D Wasserstein distance is equivalent to the Euclidean distance in the \(m_0\)-dimensional convex set \(\mathcal{T}_0\). Thus,  the Wasserstein distances satisfy the \(l_r\)-norm inequality for Euclidean distances in a convex set  \cite{boyd.2004}.
\end{proof}

\begin{figure}[t!]
	\centering
	{\includegraphics[width=0.7\linewidth]{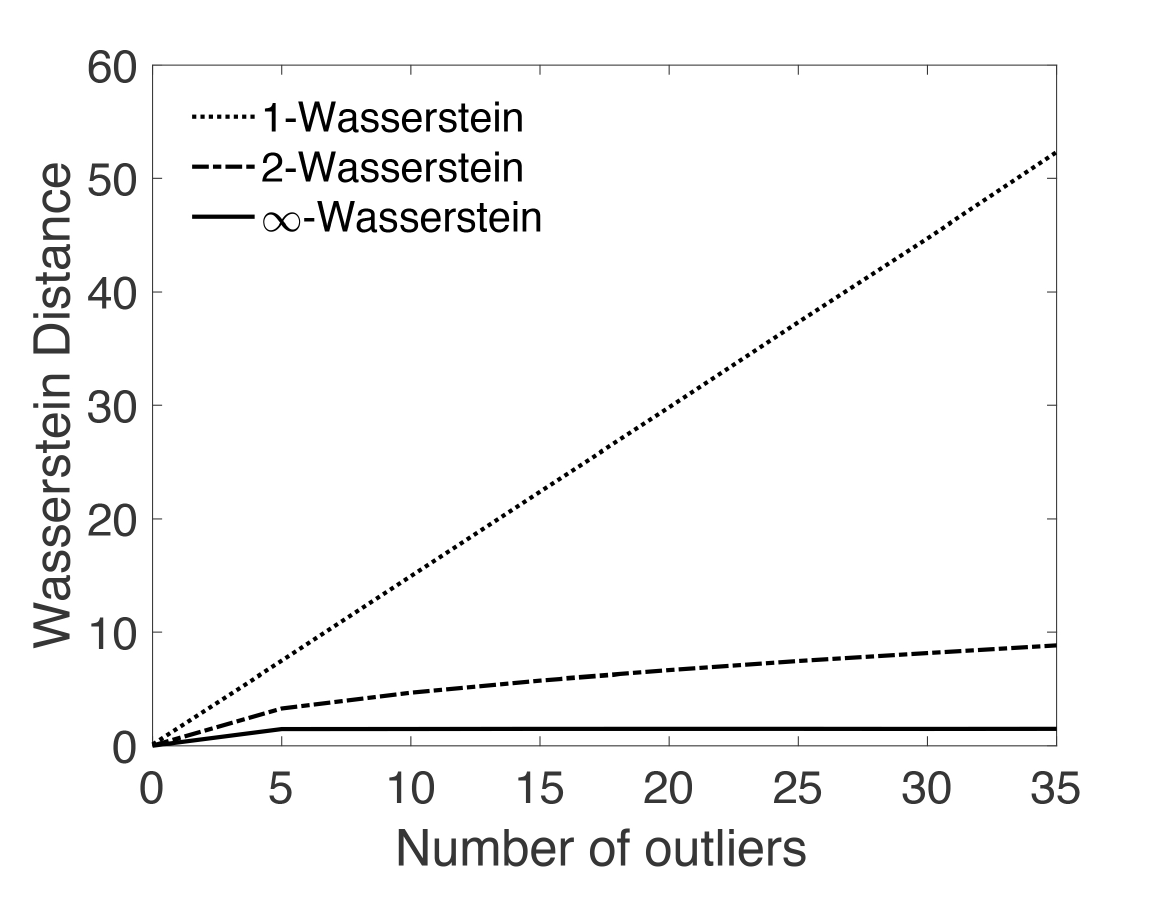}}
	\caption{Comparison of 1-, 2-, and $\infty$-Wasserstein distances as the number of outliers varies. The dataset consists of two sets of edge weight vectors of length 100, generated from $N(0, 0.01^2)$. Outliers were introduced into one set by replacing entries with values from $N(1.5, 0.01^2)$. The \(1\)-Wasserstein distance  is the upper bound of all other $r$-Wasserstein distances and it is severely impacted by the presence of outliers.}
\label{fig:out}
\end{figure}

Since the \(1\)-Wasserstein distance sets the upper bound, it is generally not the most effective metric, especially when the data contains outliers. This is illustrated through a simulation study, where two sets of edge weight vectors of length \(100\) were generated from a normal distribution \(N(0, 0.01^2)\)  (Figure \ref{fig:out}). Outliers were introduced into one of the edge weight vectors by replacing some entries with values generated from \(N(1.5, 0.01^2)\). We computed the \(1\)-, \(2\)-, and \(\infty\)-Wasserstein distances between the two sets and plotted the resulting distances in Figure~\ref{fig:out}. The distance  should be close to zero when no outliers are present. However, our observations indicate that the \(1\)-Wasserstein distance is severely impacted by the presence of outliers, showing a substantial inflation in value. On the other hand, the \(2\)- and \(\infty\)-Wasserstein distances are significantly more robust against such anomalies. Even when up to 35\% of the data consists of outliers, the \(\infty\)-Wasserstein distance provides reasonable results,
while the \(1\)-Wasserstein distance is not advisable for most applications.

\subsection{Dynamic Embedding}

\begin{figure}[t]
	\centering
	{\includegraphics[width=1\linewidth]{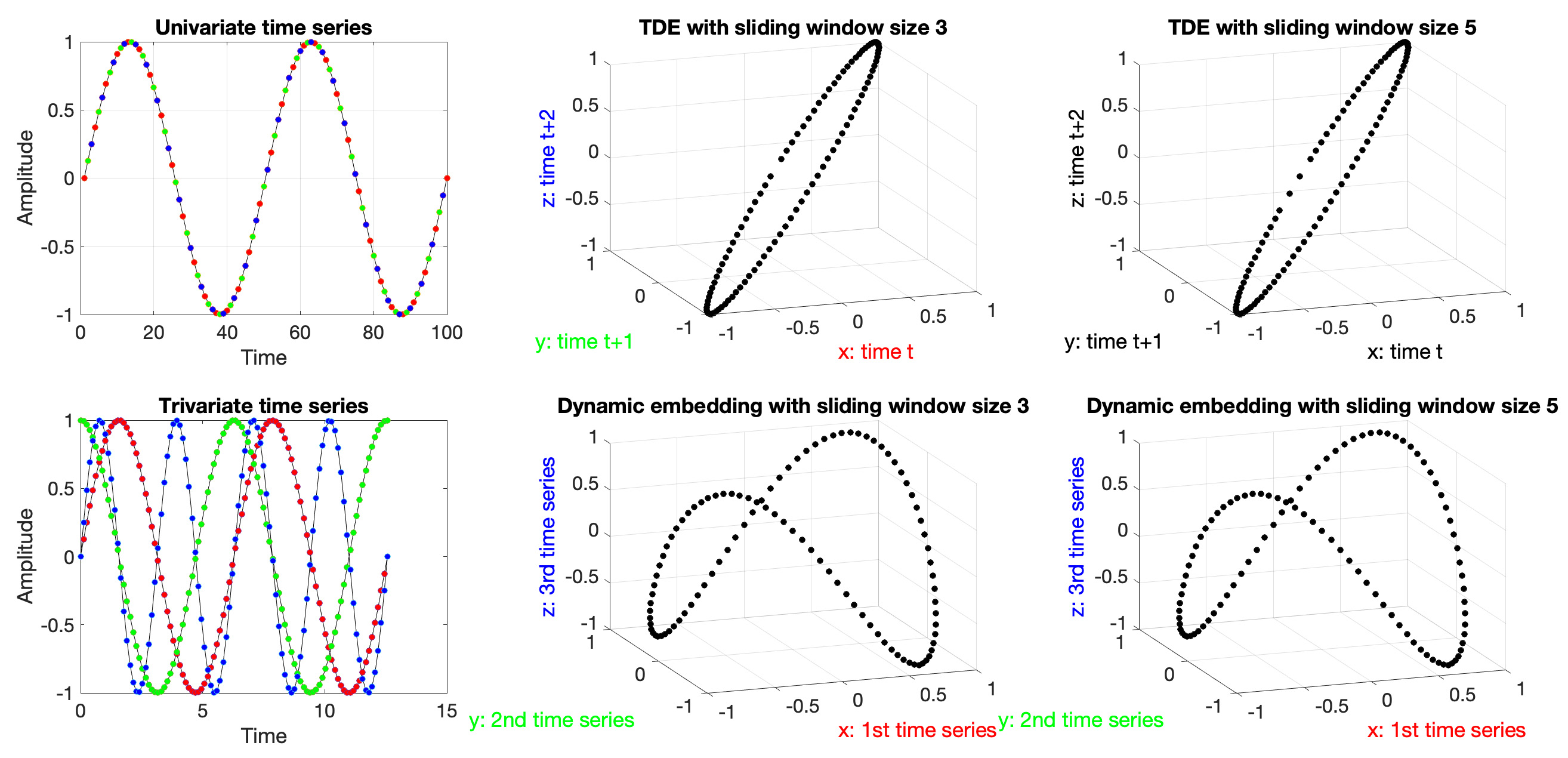}}
	\caption{Top: time delay embedding (TDE) of a univariate time series. Sliding windows of sizes \(3\) and \(5\) are used to embed the time series as a  static point cloud in \(\mathbb{R}^3\) and \(\mathbb{R}^5\), respectively. In the case of a sliding window of size \(3\), the times \(1, 4, 7, \ldots\) provide the \(x\)-coordinates (\red{red}), \(2, 5, 8, \ldots\) provide the \(y\)-coordinates (\green{green}), and \(3, 6, 9, \ldots\) provide the \(z\)-coordinates (\blue{blue}). In TDE, the window size determines the embedding dimension. Bottom: The proposed Dynamic Embedding of a trivariate time series. Sliding windows of sizes \(3\) and \(5\) are used to embed the time series into dynamically changing point clouds in \(\mathbb{R}^3\). However, due to the periodicity of the trivariate time series, the dynamically changing point clouds forms a closed loop. Unlike TDE, where the window size determines the embedding dimension, in Dynamic Embedding, it is the number of time series variables that determines the embedding dimension. Both embedding techniques are capable of capturing the underlying circular topology over time.}
\label{fig:tde_schematic}
\end{figure}

Dynamically changing functional brain imaging data can be modeled as a \(d\)-variate time series, where \(d\) is the number of brain regions under investigation. A conventional technique for transforming univariate time series data into graph-based representations employs \emph{time-delay embedding} (TDE), which is grounded in Takens' theorem \cite{seversky.2016}. Initially, TDE was aimed at uncovering the underlying dynamics of time series \cite{takens.1981}. Recently, it has been used to analyze human speech \cite{brown.2009} and to classify time series data \cite{perea.2015, emrani.2014}. In TDE, each of the \(d\) univariate time series is transformed \emph{individually} using a sliding window of size \(M\). This produces \(d\) distinct point clouds in \(\mathbb{R}^{M}\) (Figure \ref{fig:tde_schematic}). However, generalizing TDE to handle multivariate time series, such as those encountered in functional brain imaging data, is not straightforward. Moreover, TDE requires an additional step to estimate the dimension \(M\) from the data \cite{song.2020.ISBI}.
An incorrect estimation of $M$ can lead to either overfitting or underfitting. Overfitting occurs when $M$ is too large, where noise in the data is mistaken as though it was a feature. Underfitting takes place when $M$ is too small to fully capture the dynamics. Motivated by TDE, we propose a new {\em Dynamic Embedding} that transforms a $d$-variate time series ${\bf x}_t$ into a time-varying sequence of point clouds in $\mathbb{R}^{d}$.

Let \(\boldsymbol{x}_t = (x_{1t}, \ldots, x_{dt})^\top\) be a \(d\)-variate time series. At each time \(t\),
we project \(\boldsymbol{x}_t\) as a point in \(\mathbb{R}^d\), where \(x_{it}\) serves as the \(i\)-th coordinate of the point. A sliding window \([t, t+p-1]\) covering the time series \(\boldsymbol{x}_t, \ldots, \boldsymbol{x}_{t+p-1}\) yields \(p\) points in \(\mathbb{R}^d\)  (Figure \ref{fig:tde_schematic}). The Euclidean distance between these points will be used as
edge weights for building time-varying graph \(G_t = (V_t, W_t)\), where the vertex set \(V_t = \{ \boldsymbol{x}_t, \ldots, \boldsymbol{x}_{t+p-1} \}\) and the edge weights \(W_t\) are the Euclidean distances between corresponding points. Subsequently, we construct graph filtrations to obtain dynamically changing 0-dimensional and 1-dimensional persistent diagrams, represented as time varying birth and death values
$$B(G_t) = \{ b_{1t}, \cdots, b_{m_{0t}} \}, \quad  D(G_t) = \{ d_{1t}, \cdots, d_{m_{1t}} \}$$
respectively. Unlike TDE, which implicitly incorporates the dynamic features of the time series into a point cloud,  Dynamic Embedding has the distinct advantage of easily extracting the dynamic topological characteristics of the time series.

Figure \ref{fig:tde_schematic} illustrates the difference between TDE and Dynamic Embedding. A univariate time series is constructed as
\[
x_t = \sin(4\pi t), \quad t  = 0, 1, 2, \cdots, 100,
\]
and used in TDE with window sizes \(M = 3, 5\).
For the window size \(M = 3\), TDE produces a static point cloud in \(\mathbb{R}^3\). The coordinates are color-coded: red for the \(x\)-coordinate, green for the \(y\)-coordinate, and blue for the \(z\)-coordinate. TDE reveals a cyclic pattern due to the periodicity of the time series. A trivariate time series is constructed as
\[
{\bf x}_t = (\sin(4\pi t), \cos(4\pi t), \sin(8\pi t))^{\top}, \quad t = 0, 1, 2, \cdots, 100.
\]

Dynamic Embedding is applied to produce a dynamically changing point cloud in \(\mathbb{R}^3\). Due to the periodicity, it warps around to form a closed loop. The coordinates are color-coded as red (\(x\)-coordinate), green (\(y\)-coordinate), and blue (\(z\)-coordinate), each corresponding to one of the time series. The Dynamic Embedding also reveals the cyclic pattern resulting from the periodicity of the time series.

\begin{figure}[t]
	\centering
	{\includegraphics[width=1\linewidth]{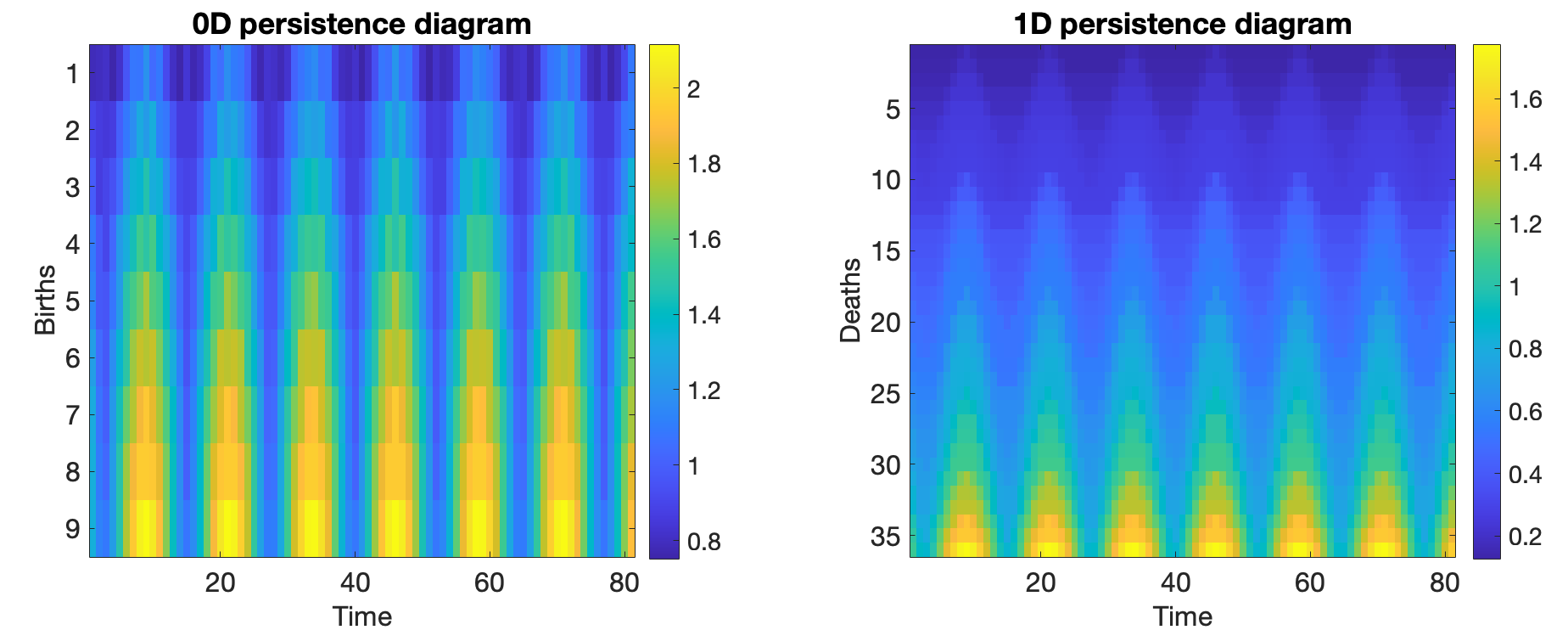}}
	\caption{The time-varying persistence diagrams for the Dynamic Embedding on the trivariate time series in Figure \ref{fig:tde_schematic}-bottom with a sliding window size of 10. The embedding yields 9 dynamically changing cloud points per window. The persistence diagrams generated from the graph filtration yield sorted birth (0-dimensional) and death (1-dimensional) values. If a smaller window size, such as 3, is used, it produces only 2 points, rendering the topology too crude for accurate estimation. A larger window size was selected to provide sufficient points for a more accurate estimation of the underlying topology. The periodic pattern observed in the persistence diagrams is a consequence of the periodicity present in the trivariate time series.}
\label{fig:tde_dynamic}
\end{figure}

Consider two \(d\)-variate time series \({\bf x}_t=(x_{1t}, \ldots, x_{dt})^\top\) and \({\bf y}_t=(y_{1t}, \ldots, y_{dt})^\top\) \cite{kakizawa1998discrimination, shumway2003time, fryzlewicz2009consistent, bohm2010classification}. We aim to determine the topological differences between these time series. Following Dynamic Embedding, we construct time-varying graphs \(X_t\) and \(Y_t\) with \(p\) nodes corresponding to the time series \({\bf x}_t\) and \({\bf y}_t\), respectively. The total number of birth values of connected components and the total number of death values of cycles are given by \cite{song.2023}
\[
m_0 = p-1, \quad m_1 = 1+ \frac{p(p-3)}{2}.
\]

We can compute the 0-dimensional and 1-dimensional persistence diagrams at time \(t\) corresponding to the time series \({\bf x}_t\) and \({\bf y}_t\). The \(r\)-Wasserstein distance between the 0-dimensional persistence diagrams at time \(t\) is given by
\[
\mathfrak{L}_{r}^b({\bf x}_t, {\bf y}_t) = \left(\sum_{k=1}^{m_0} \left| b_{kt}^{{\bf x}} - b_{kt}^{{\bf y}} \right|^r\right)^{1/r},
\]
where \(b_{kt}^{{\bf x}}\) and \(b_{kt}^{{\bf y}}\) are the \(k\)-th smallest birth values associated with the graphs \(X_t\) and \(Y_t\), respectively. Similarly, the \(r\)-Wasserstein distance between the 1-dimensional persistence diagrams at time \(t\) is
\[
\mathfrak{L}_{r}^d({\bf x}_t, {\bf y}_t) = \left(\sum_{k=1}^{m_1} \left| d_{kt}^{{\bf x}} - d_{kt}^{{\bf y}} \right|^r\right)^{1/r},
\]
where \(d_{kt}^{{\bf x}}\) and \(d_{kt}^{{\bf y}}\) are the \(k\)-th smallest death values associated with the graphs \(X_t\) and \(Y_t\), respectively. Lastly, the \(\infty\)-Wasserstein distances are given by
\[
\mathfrak{L}_{\infty}^b({\bf x}_t, {\bf y}_t) = \max_{1\leq k \leq m_0} \left| b_{kt}^{{\bf x}} - b_{kt}^{{\bf y}} \right|
\]
and
\[
\mathfrak{L}_{\infty}^d({\bf x}_t, {\bf y}_t) = \max_{1\leq k \leq m_1} \left| d_{kt}^{{\bf x}} - d_{kt}^{{\bf y}} \right|.
\]

Subsequently, we propose to combine the $r$-Wasserstein distances over all time as
\begin{align*}
\mathcal{L}_r({\bf x}_t, {\bf y}_t)
= \sum_t \mathfrak{L}_{r}^b({\bf x}_t, {\bf y}_t)  + \sum_t \mathfrak{L}_{r}^d({\bf x}_t, {\bf y}_t).
\end{align*}

For $r\to \infty$, we have
\begin{align*}
\mathcal{L}_\infty({\bf x}_t, {\bf y}_t)
=\sum \limits_{t} \mathfrak{L}_{\infty}^b({\bf x}_t, {\bf y}_t)    + \sum \limits_{t} \mathfrak{L}_{\infty}^d({\bf x}_t, {\bf y}_t)  .
\end{align*}

A large value of \(\mathcal{L}_r({\bf x}, {\bf y})\) or  \(\mathcal{L}_\infty({\bf x}, {\bf y})\) suggests that the dynamic persistence diagrams generated from the two time series \({\bf x}_t\) and \({\bf y}_t\) are \emph{topologically distant}. This indicates a significant difference in the 0-dimensional and 1-dimensional topological features between \({\bf x}_t\) and \({\bf y}_t\). Since the distributions of the proposed test statisticss \(\mathcal{L}_r({\bf x}_t, {\bf y}_t)\) and \(\mathcal{L}_\infty({\bf x}_t, {\bf y}_t)\) are unknown, we employ the permutation test for inference. The permutation test does not make any assumptions about the underlying probability distribution \cite{chung.2019.CNI,thompson.2001, zalesky.2010, nichols.2002, winkler.2016}.

\subsection{Topological inference on Dynamic Embedding}

Suppose we have \(m\) and \(n\) \(d\)-variate time series in two  groups. Given two sets of \(d\)-variate time series, \(\{ {\bf x}^1, {\bf x}^2, \ldots, {\bf x}^m  \}\) and \(\{  {\bf y}^1, {\bf y}^2, \ldots, {\bf y}^n  \}\), we aim to compute the $p$-value to test the null hypothesis that the two sets of time series are topologically equivalent. The topological similarity is measured through the Wasserstein distance. For each pair of time series \( {\bf x}^i \) and \( {\bf y}^j \), we compute the Wasserstein distance \( \mathcal{L}_r({\bf x}^i, {\bf y}^j) \) using birth and death values obtained from topological features.

We consider the combined ordered set
\[
{\bf z} = \{ {\bf z}^1, \cdots, {\bf z}^{m+n} \} = \{  {\bf x}^1, {\bf x}^2, \ldots, {\bf x}^m, {\bf y}^1, {\bf y}^2, \ldots, {\bf y}^n  \}.
\]

The between-group distance \( d({\bf z}) \) is given by the total sum of pairwise Wasserstein distances:
\[
  d({\bf z})  = \sum_{i=1}^{m} \sum_{j=1}^{n} \mathcal{L}_r({\bf x}^i, {\bf y}^j).
\]

This can be used as a test statistic. Since the distribution is unknown, the statistical significance can be determined through the scalable version of permutation test through transpositions \cite{chung.2019.CNI,song.2023}. We first compute the pairwise distance matrix \( D = (d_{ij}) = ( \mathcal{L}_r({\bf z}^i, {\bf z}^j) ) \). Once \(D\) is computed, there is no need to recompute it for each permutation. We only need to shuffle the entries of \(D\) by permutation as follows.

The between-group distance can be expressed as
\[
d({\bf z}) = \sum_{i=1}^m \sum_{j=m+1}^{m+n} d_{ij} = {\bf 1}_m^\top D {\bf 1}_n,
\]
where where \( {\bf 1}_m \) and \( {\bf 1}_n \) are indicator vectors of dimension \( m+n \) for each group such that
$${\bf 1}_m = [1, \cdots, 1, 0, \cdots, 0]^{\top}, \quad {\bf 1}_n = [0, \cdots, 0, 1, \cdots, 1]^{\top}.$$

There are $m$ 1's in ${\bf 1}_m$ and $n$ 1's in ${\bf 1}_n$.
Let \(\pi \in \mathbb{S}_{m+n} \) be a permutation for integers $\{1, \cdots, m+n \}$ in the permutation group of order $m+n$.
For permutation \( {\bf z}^{\pi} = \{ {\bf z}^{\pi(1)}, \cdots, {\bf z}^{\pi(m+n)} \}\) which permutes entries of ${\bf z}$, the permuted between-group distance \(d({\bf z}^\pi)\) can be represented as
\bqn d({\bf z}^\pi) = {\bf 1}_m^\top \Pi D \Pi ^\top {\bf 1}_n, \label{eq:dzpi} \eqn
with the permutation matrix \( \Pi= (q_{ij}) \) defined as a \( (m+n) \times (m+n) \) matrix such that \( q_{ij} = 1 \) if \( \pi(i) = j \) and \( q_{ij} = 0 \) otherwise. After the permutation, the Wasserstein distance incrementally changes to
\bqn  d({\bf z}^{\pi}) = d({\bf z}) + \Delta d, \label{eq:d4} \eqn
where $\Delta d = {\bf 1}_m^\top (\Pi D \Pi^\top - \Pi ) {\bf 1}_n$
is the increment over permutation $\pi$. The computation of \( d({\bf z}^{\pi}) \) leverages term $d({\bf z})$ recycled from the previous iteration. When only one entry from each group is permuted, we have transpositions, and the iterative update for \( \Delta d \) in equation~\eqref{eq:d4} involves only a small number of terms and we can further increase the rate of convergence \cite{chung.2019.CNI,song.2023}. Thus, in numerical implementations, we intersperse a full permutation after every 1000 transpositions.

Since $d({\bf z})$ increases as the number of networks $m$ and $n$ increases, we normalized it as
$$\lambda ({\bf z}) = \frac{d({\bf z})}{ \sum_{i=1}^{m+n} \sum_{j=1}^{m+n} \mathcal{L}_r ({\bf z}^i, {\bf z}^j)- d({\bf z})}.$$

The sum of all pairwise distances $\sum_{i=1}^{m+n} \sum_{j=1}^{m+n} \mathcal{L}_r$ is fixed regardless of how we assign group labels. The denominator is the within-group distance, which is the sum of all pairwise distance within each group. Subsequently, we use the ratio $\lambda ({\bf z})$ as a test statistic for testing the equality of two sets of $d$-variate time series. The $p$-value is then calculated as the proportion of permutations where \(\lambda ({\bf z}^\pi) \) exceeds \( \lambda ({\bf z}) \):
\bqn p\mbox{-value} =\frac{1}{(m+n)!} \sum_{\pi \in \mathbb{S}_{m+n}  } \mathcal{I} \Big(  \lambda({\bf z}^{\pi})  > \lambda({\bf z})  \Big),  \label{eq:pvalue} \eqn
where $\mathcal{I}$ is an indicator variable taking value 1 if the argument is true and 0 otherwise \cite{chung.2019.CNI}. Since the computation for every possible permutation is too time consuming, we usually perform uniform sampling in $\mathbb{S}_{m+n}$. Once incremental formula (\ref{eq:d4}) is identified, the $p$-value can be computed iteratively over permutations.  At the $k$-th permutation,  the $p$-value $p_k$ is updated as
\bqn (k+1) p_{k+1} = kp_k  +
 \mathcal{I} \Big(  \lambda ({\bf z}^{\pi})  > \lambda ({\bf z})  \Big).\label{eq:pk} \eqn

The MATLAB code for Dynamic Embedding and topological inference are available at \url{https://github.com/laplcebeltrami/PH-STAT}.

\section{Simulation studies}

\begin{figure}[t]
	\centering
	\includegraphics[width=1\linewidth]{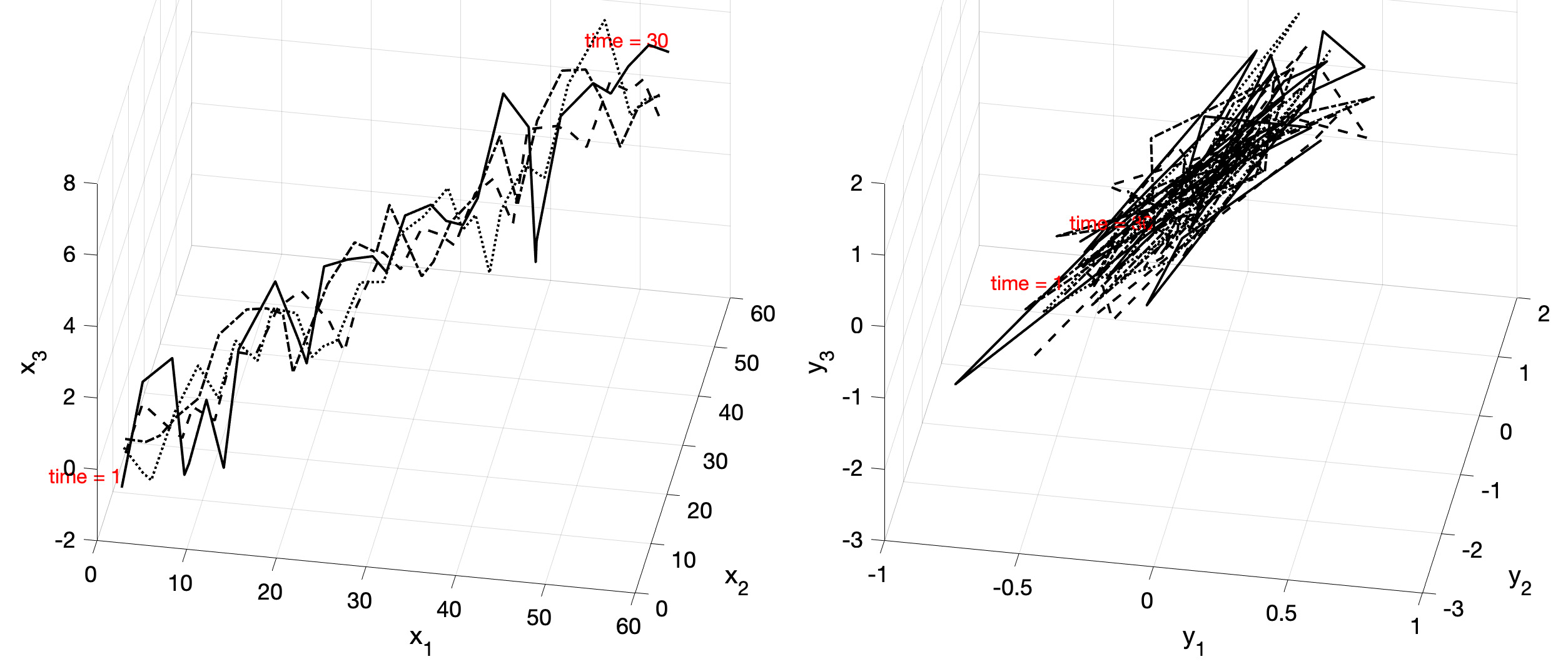}
	\caption{Four independently simulated trivariate time series \({\bf x}_t = (x_1(t), x_2(t), x_3(t))^{\top} \) (equation \ref{eq:study-1})  and \({\bf y}_t = (y_1(t), y_2(t), y_3(t))^{\top} \)  (equation \ref{eq:study-2}) in Study 1. Each trivariate time series is represented as a random walk of length 30 in \(\mathbb{R}^3\). The  Wasserstein distance was able to discriminate between samples generated from \({\bf x}_t\) and \({\bf y}_t\), but did not detect any differences in samples generated solely from \({\bf x}_t\).}
\label{fig:simulation1}
\end{figure}

Since there is no ground truth in real data, we performed two simulation studies with specified ground truths for testing false-negatives and false-positives. These controlled experiments  were utilized to assess the performance of the proposed method. Consider a trivariate time series generated by a vector autoregressive model of order one \cite{zivot2006vector, lutkepohl2013vector, haslbeck2021tutorial}:
\bqn
{\bf x}_t = \mathbf{a} + \mathbf{b}t + \Theta{\bf x}_{t-1} + \p{\xi}_t,
\label{eq:study-1}
\eqn
with noise $\p{\xi}_t \sim N (0, \Sigma)$ at time \(t\) (Figure \ref{fig:simulation1}).
We also considered a different trivariate time series consisting solely of Gaussian white noise at time $t$:
\bqn
{\bf y}_t \sim N (0, \Sigma).
\label{eq:study-2}
\eqn

The trend parameters were set to $\mathbf{a} = (1, 1, 0)^\top$ and $\mathbf{b} = (1.5, 2, 0)^\top$ for  \(1 \leq t \leq 30\).
The coefficient and covariance matrices are chosen as
\begin{align*}
\Theta = \begin{bmatrix}
0.2 & -0.1 & 0.5 \\
-0.4 & 0.2 & 0 \\
-0.1 & 0.2 & 0.3
\end{bmatrix},
\quad \Sigma = \begin{bmatrix}
0.1 & 0.01 & 0.3 \\
0.01 & 0.5 & 0 \\
0.3 & 0 & 1
\end{bmatrix}.
\end{align*}

For simulations, we generated 10 time series for each time series ${\bf x}_t, {\bf y}_t$ and ${\bf z}_t$. Used $p=10$  as size of sliding window for Dynamic Embedding. The transposition test with 100,000 transpositions were used to determine statistical signfance.

\subsection{When there is difference in time series}
We tested if the method can detect topological differences in time series ${\bf x}_t$ and ${\bf y}_t$ (Figure \ref{fig:simulation1}). We used \(\mathcal{L}_2(\mathbf{x}_t, \mathbf{y}_t)\) as the test statistic. Each simulation is independently replicated 10 times, and the resulting $p$-values are averaged. The computed $p$-values are $(0.50 \pm 0.97) \times 10^{-5}$, signifying a robust topological disparity between the two trivariate time series.

\subsection{When there is no difference in time series}
To assess the false-positive rate of our method, we introduce another time series \(\mathbf{z}_t\), which is modeled identically to \(\mathbf{x}_t\) but with independently generated noise (Figure \ref{fig:simulation1}). In this case, we do not expect our method to identify any topological differences between \(\mathbf{x}_t\) and \(\mathbf{z}_t\). The simulation was conducted 10 times independently, and the results were averaged. The computed $p$-values are \(0.33 \pm 0.20\), confirming that our method does not incorrectly identify topological differences.

\begin{figure}[t]
	\centering
	\includegraphics[width=1\linewidth]{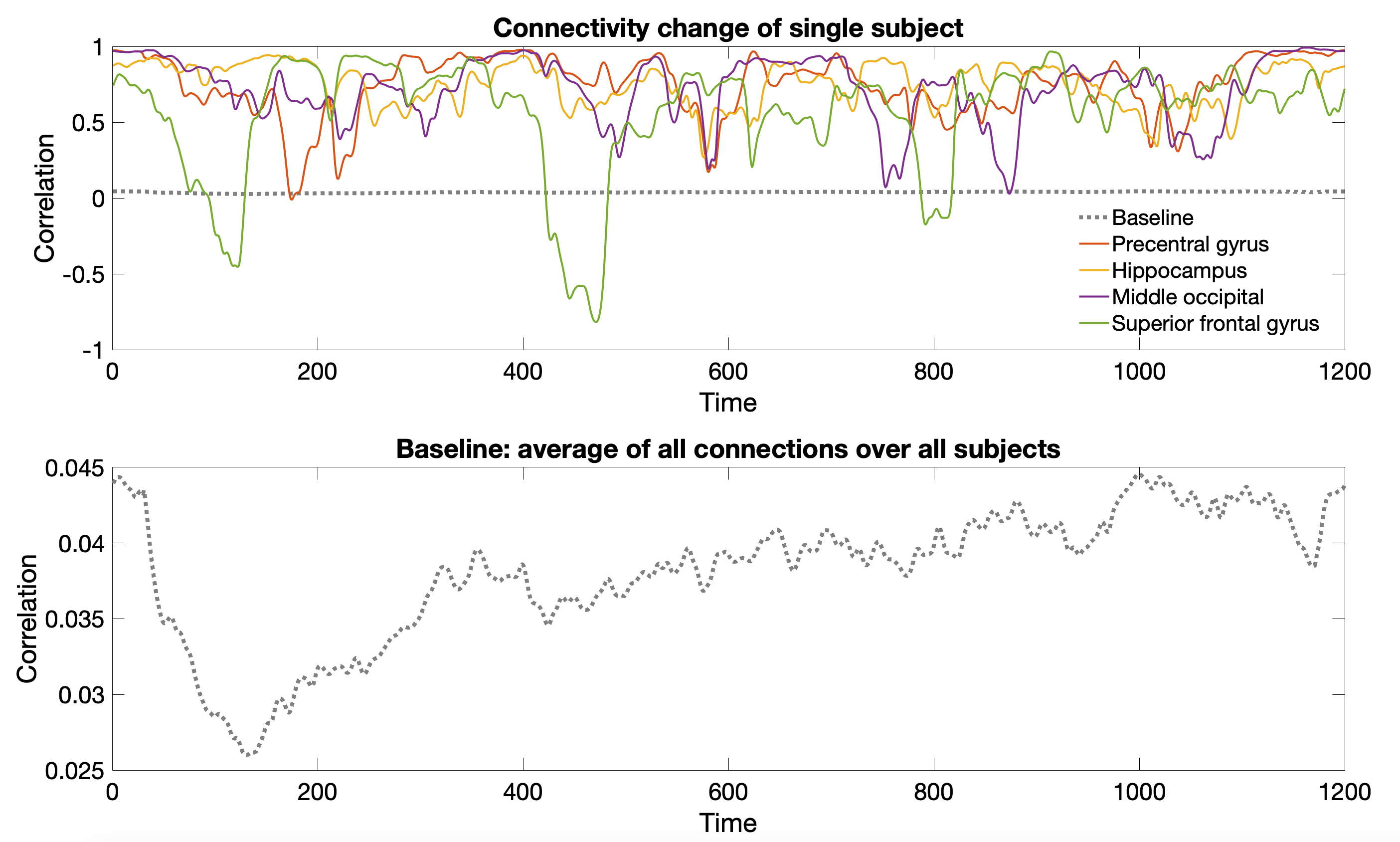}
	\caption{Top: Time series of functional brain connectivity between the left and right hemispheres across 4 regions. We analyzed a total of 6670 time series of brain connectivity between 116 regions in this study. Bottom: The average time series of functional brain connectivity, averaged over all connections and subjects. We removed the first 60 time points in the subsequent analysis. Although a slight increasing trend is observed, it is negligible when compared to the individual connectivities.}
	\label{fig:AAL}
\end{figure}

\section{Application to functional brain images} \label{real}
\subsection{Data description} We used  a resting-state fMRI dataset collected as part of the Human Connectome Project (HCP) \cite{vanessen.2012, glasser2013minimal}. The dataset consists of  fMRI scans of $412$ subjects ($172$ males and $240$ females) measured over $1200$ time points using a gradient-echoplanar imaging sequence \cite{song.2023, anand.2023.TMI}. Since there are more than $300000$ voxels in an fMRI scan, we parcellated the brain into 116 non-overlapping anatomical regions using the Automated Anatomical Labeling (AAL) template and averaged rsfMRI over each parcellation \cite{tzourio.2002,desikan.2006, hagmann.2007, arslan.2018}.  The head movements cause serious spatial artifacts in functional connectivity \cite{power.2012, van.2012, satterthwaite.2012, caballero.2017}. We calculated the framewise displacements from the three translational displacements and three rotational displacements at each time point to measure the head movement from one volume to the next. The volumes with framewise displacement larger than $0.5$ mm from their neighbors (one back and two forward time points) were scrubbed \cite{van.2012, power.2012, huang.2020.NM}. More than one third of $1200$ volumes were scrubbed in $12$ subjects. Thus, we  removed the 12 subjects ($4$ males and $8$ females) and used 400 subjects for the study.  Following  \cite{allen.2014}, we adopted window size 60 repetition time (TR), which is time taken for the MRI sequence to acquire one whole volume. Since TR=0.72 seconds in HCP, it is 43.2 seconds in physical time. Subsequently we computed the Pearson correlation between brain regions in each sliding window in measuring the strength of time varying brain connectivity.  The details on the rsfMRI preprocessing we performed is explained in our previous study \cite{huang.2020.NM}. There are total $(116 \times 115)/2 = 6670$ time series of brain connectivity between 116 regions per subject that is used as an input to this study.

\begin{figure}[t]
	\centering
	\includegraphics[width=1\linewidth]{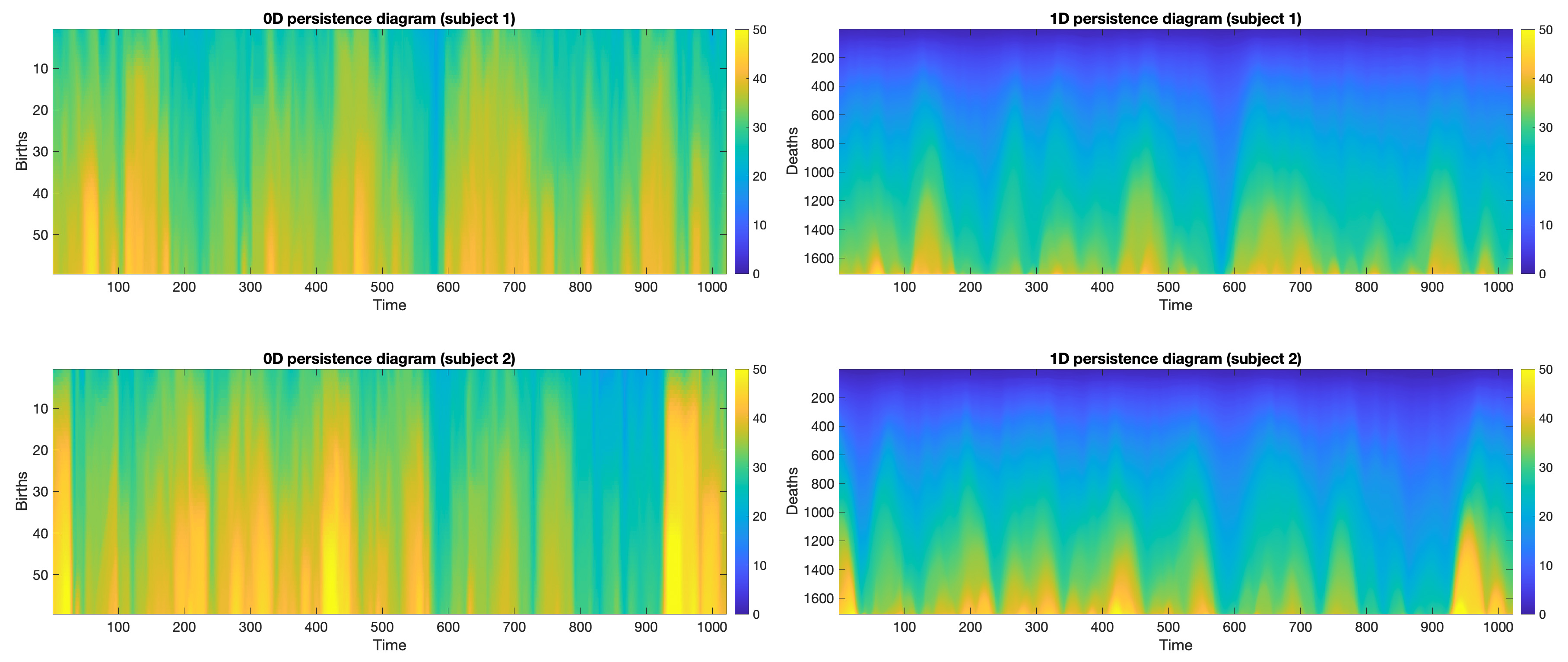}
	\caption{Dynamically changing persistence diagrams for subject 1 (top) and subject 2 (bottom). The persistence diagrams were computed on time varying graphs obtained through Dynamic Embedding in $\mathbb{R}^{60}$. We employed the 2-Wasserstein distance to measure topological similarity between the subjects within a window of 60 TRs, equivalent to 43.2 seconds.}
	\label{fig:PD-dynam}
\end{figure}

The time series are averaged over all connections to determine if there is an overall trend (Figure \ref{fig:AAL}-bottom). We observe a sharp decline in the initial measurements of fMRI, which is usually considered as artifacts included by the large variance at the beginning of each MRI scan \cite{huang.2020.NM,diedrichsen.2005}. Thus, we removed the first $60$ time points and only used the remaining $1140$ time points in the subsequent analysis. There exists a slow increasing trend (Figure~\ref{fig:AAL}-bottom). However, the trend is  negligible when compared to the individual connectivities. Further, the time-varying graph $G_t$ obtained in the Dynamic Embedding is invariant of changes in the baseline fMRI signal. If there's a general upward or downward trend in the fMRI data over time, this will not have an impact on the structure of graph $G_t$. Therefore, we applied our method directly to the fMRI connectivity without the need for detrending.

\subsection{Global topological inference} We applied our method to determine if there is overall topological difference between resting-state functional brain connectivity between males and females. Sex differences in rs-fMRI have been a subject of considerable research interest. Studies have found that males and females exhibit different patterns of functional connectivity in the default mode network \cite{biswal.2010}. Research on specific brain regions like the amygdala has shown sex-specific connectivity patterns that could be linked to emotional processing \cite{kilpatrick.2006}. Sex differences in the rs-fMRI connectivity of the brain could be related to cognitive performance and behavioral traits \cite{satterthwaite.2015}. In diffusion tensor imaging study, males have been observed to have greater within-hemisphere connectivity, whereas females have greater between-hemisphere connectivity \cite{ingalhalikar.2014}.

\begin{figure}[!t]
	\centering
	\includegraphics[width=1\linewidth]{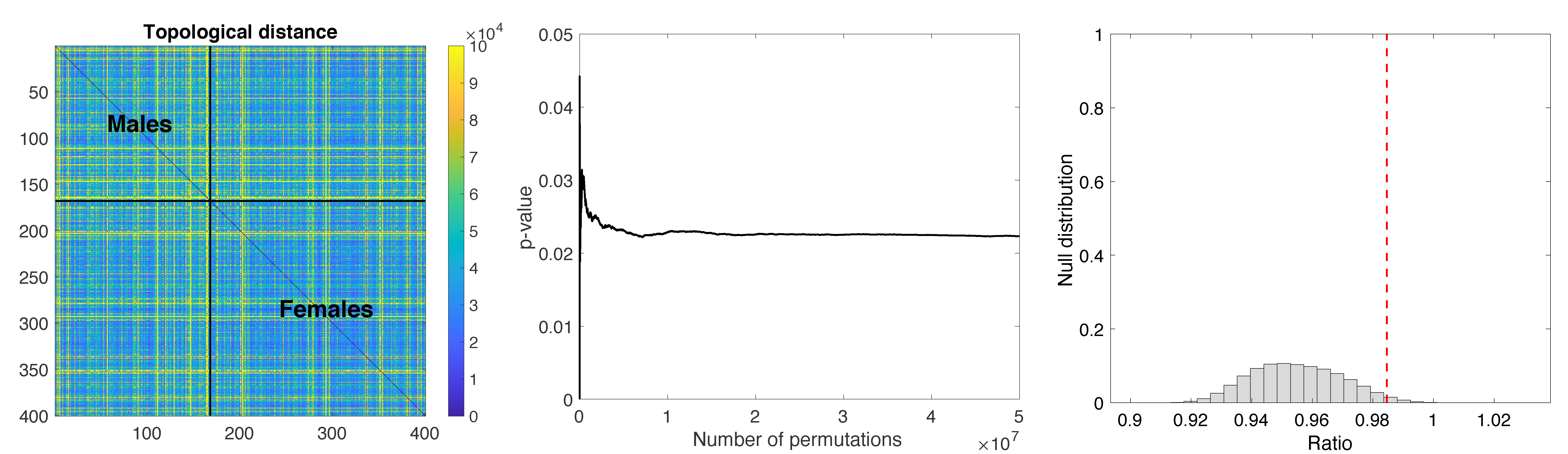} 
	\caption{Left: Distance matrix $\mathcal{L}_2 ({\bf z}^i, {\bf z}^j)$. Males have slightly larger topological distance compared to females. Middle: plot showing the convergence of $p$--value over increasing permutations. At 50 million permutations, the estimated $p$-value is stable. Right: the empirical null distribution obtained from 50 million permutations. The red line is the observed ratio
	statistic of the between-group over within-group distances.}
\label{fig:real_hist}
\end{figure}

We performed Dynamic Embedding on \(d=6670\) time-varying connectivities. We employed a sliding window approach with a window size of \(p=60\). This window size was chosen to match the sliding window used for computing correlations to ensure consistency across different analytical methods and to minimize aliasing artifacts. Aliasing occurs when a signal is sampled at a rate leading to distortions or inaccuracies in the reconstructed signal. By aligning the window sizes, we enhanced the robustness and interpretability of our results. Figure \ref{fig:PD-dynam} displays the time-varying persistence diagrams for two subjects. There are total $m=168$ males and $n=232$ females. We performed the permutation test with 50 million random transpositions intermixed with one full permutation every 1000 transpositions on the Wasserstein distance  $\mathcal{L}_2$ in discriminating sex. We obtain the observed ratio  statistic of $0.9846$ and the corresponding $p$-value $0.0223$. Thus, we conclude that the proposed method distinctly discriminates the functional connectivity between females and males. Figure~\ref{fig:real_hist} plots the histogram of the test statistic and the corresponding observed test statistic (in dashed red).

\section{Conclusion}
Topological data analysis (TDA) based approaches have been previously applied to static brain imaging studies \cite{chung.2019.ISBI, bassett.2017, wang.2018.annals}, focusing on identifying topological features that characterize brain signals. However, these methods are primarily tailored for investigating static summaries of brain networks, which inherently limits their ability to provide comprehensive insights. In this paper, we introduce a new framework for dynamic-TDA that is capable of statistically discriminating between two groups of multivariate time series. Our proposed methodology builds on persistent homology over multivariate time series and constructs a statistical measure based on $r$-Wasserstein distances between persistence diagrams. We have applied this approach to resting-state fMRI data from both female and male human subjects and discovered significant topological differences in terms of combined 0D and 1D topological distances between the two groups. Given the success of the proposed approach in differentiating multivariate time series, we believe that it can be extended to analyze other types of time series data as well.

\section*{Acknowledgments}
This study is funded by NIH EB02875, MH133614, NSF MDS-2010778 and the King Abdullah University of
Science and Technology \break(KAUST) CRG. The use of CHAT-GPT is acknowledged.


\providecommand{\href}[2]{#2}
\providecommand{\arxiv}[1]{\href{http://arxiv.org/abs/#1}{arXiv:#1}}
\providecommand{\url}[1]{\texttt{#1}}
\providecommand{\urlprefix}{URL }

\medskip
Received October 2023; revised October 2023; early access December 2023.
\medskip

\end{document}